\documentclass[preprint,amsmath,amssymb]{revtex4-1}

\usepackage{graphicx}
\usepackage{dcolumn}
\usepackage{bm}
\usepackage{caption}
\begin{document}

\title{Evidence for hydrogen two-level systems in atomic layer deposition oxides}

\author{M. S. Khalil}
\email{moe@lps.umd.edu}
\affiliation{ 
Laboratory for Physical Sciences, College Park, MD, 20740
}
\affiliation{ 
Department of Physics, University of Maryland, College Park, MD, 20742
}

\author{M. J. A. Stoutimore}%
\affiliation{ 
Laboratory for Physical Sciences, College Park, MD, 20740
}
\affiliation{ 
Department of Physics, University of Maryland, College Park, MD, 20742
}

\author{S. Gladchenko}%
\affiliation{ 
Laboratory for Physical Sciences, College Park, MD, 20740
}
\affiliation{ 
Department of Physics, University of Maryland, College Park, MD, 20742
}

\author{A. M. Holder}
\affiliation{ 
 Department of Chemistry and Biochemistry, Department of Chemical and Biological Engineering, University of Colorado, Boulder, CO, 80309}

\author{C. B. Musgrave}
\affiliation{ 
 Department of Chemistry and Biochemistry, Department of Chemical and Biological Engineering, University of Colorado, Boulder, CO, 80309}

\author{A. C. Kozen}
\affiliation{ 
Department of Material Science, University of Maryland, College Park, MD, 20742}

\author{G. Rubloff}
\affiliation{ 
Department of Material Science, University of Maryland, College Park, MD, 20742}

\author{Y. Q. Liu}
\affiliation{ 
Department of Chemistry and Chemical Biology, Harvard University, Cambridge, MA, 02138}

\author{R. G. Gordon}
\affiliation{ 
Department of Chemistry and Chemical Biology, Harvard University, Cambridge, MA, 02138}

\author{J. H. Yum}
\affiliation{ 
Department of Electrical and Computer Engineering, University of Texas, Austin, TX, 78758}

\author{S. K. Banerjee}
\affiliation{ 
Department of Electrical and Computer Engineering, University of Texas, Austin, TX, 78758}

\author{C. J. Lobb}
\affiliation{ 
Department of Physics, University of Maryland, College Park, MD, 20742
}
\affiliation{ 
Joint Quantum Institute and Center for Nanophysics and Advanced Materials, University of Maryland, College Park, MD, 20742
}

\author{K. D. Osborn}%
 \email{osborn@lps.umd.edu}
\affiliation{ 
Laboratory for Physical Sciences, College Park, MD, 20740
}
\date{\today}

\begin{abstract}

Two-level system (TLS) defects in dielectrics are known to limit the performance of electronic devices. We study TLS using millikelvin microwave loss measurements of three atomic layer deposited (ALD) oxide films--crystalline BeO ($\rm{c-BeO}$), amorphous $\rm{Al_2O_3}$ ($\rm{a-Al_2O_3}$), and amorphous $\rm{LaAlO_3}$ ($\rm{a-LaAlO_3}$)--and interpret them with room temperature characterization measurements. We find that the bulk loss tangent in the crystalline film is 6 times higher than in the amorphous films. In addition, its power saturation agrees with an amorphous distribution of TLS. Through a comparison of loss tangent data to secondary ion mass spectrometry (SIMS) impurity analysis we find that the dominant loss in all film types is consistent with hydrogen-based TLS. In the amorphous films excess hydrogen is found at the ambient-exposed surface, and we extract the associated hydrogen-based surface loss tangent. Data from films with a factor of 40 difference in  carbon impurities revealed that carbon is currently a negligible contributor to TLS loss. 

\end{abstract}

\maketitle 

In the standard model of amorphous materials, low-temperature dielectric and elastic properties are described by a distribution of tunneling two-level systems (TLS). \cite{Phillips1972,VONSCHICKFUS1977} Although this model has been known for decades, the microscopic nature of these tunneling states is generally unknown and only in rare circumstances are elastic or electrical properties correlated with specific impurities. 

There is renewed interest in these defects \cite{Stoutimore2012} because they have been found to decrease the quantum coherence in superconducting qubits \cite{Martinis2005} and increase the noise in astronomical photon detectors. \cite{Day2003,Gao2008} Work to decrease the deleterious effects of TLS in devices has yielded new Josephson junction barrier dielectrics, \cite{Oh2006} high-Q crystalline capacitors, \cite{Weber2011} and methods to reduce the need for dielectrics. \cite{Cicak2009,Paik2011} Atomic layer deposited (ALD) films allow unprecedented control over material thickness and in the future may allow for high quality Josephson junction barriers and thin-film capacitors for qubits. \cite{Kozen2013,Peng2007}

To better understand TLS in amorphous and crystalline oxides we measure three different ALD-grown film types: $\rm{c-BeO}$, \cite{Yum2012}$  \rm{a-Al_2O_3}$, \cite{Matero2000} and $\rm{a-LaAlO_3}$. \cite{Lim2004} While $\rm{a-Al_2O_3}$ is a prevalent thermally grown Josephson junction barrier for qubits, $\rm{BeO}$ and $\rm{LaAlO_3}$ are also candidate barriers for Josephson junctions made with ALD. The $\rm{c-BeO}$ film is grown on high-resistivity silicon and the amorphous films were grown on sapphire. Coplanar strip aluminum resonators \cite{Khalil2011} at 6.4 GHz were fabricated on top of the films by sputter deposition followed by wet etching. As with previous measurements, \cite{Khalil2012} we extract the internal quality factor $(Q_i)$ and the voltage amplitude of the resonance using a transmission measurement. The devices are mounted on the mixing chamber of a dilution refrigerator and are measured at 34-80 mK. The input microwave line has 20 dB of attenuation at the mixing chamber and 1.5 K stage such that the resonator is excited with less than a single photon of noise. The output microwave line has 36 dB and 18 dB of isolation at the mixing chamber and the 1.5 K stage, respectively, and isolates the resonator from noise on the line at 4K created at the input of a HEMT amplifier.

\begin{figure}
\includegraphics[trim = 0.1cm 0 0.5cm 0.4cm, clip]{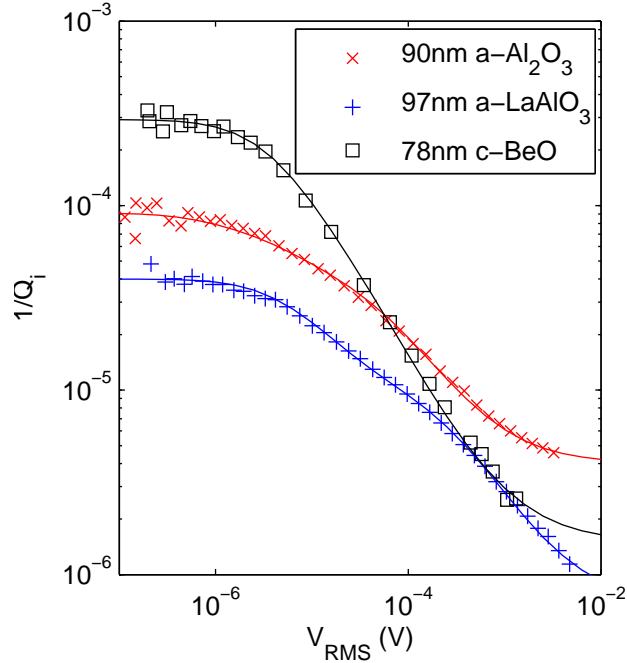}
\caption{\label{fig:mat_loss}  Inverse internal quality factor $(1/Q_i)$ of coplanar strip resonators on three film types. Fits (solid curves) are performed using the loss expression for amorphous oxides (Eq.~\ref{eq:tandForm}) with the parameters from Table~\ref{tab:FitPar}.}
\end{figure}

Figure~\ref{fig:mat_loss} shows a plot of the inverse quality factor of the coplanar strip resonators on three different films taken at 35 mK as a function of the maximum RMS voltage across the coplanar strip electrodes. This would correspond to the loss tangent of the films as a function of the electric field, $E(\vec{r})$ if the electric field energy were uniform and entirely within the lossy dielectric as with a parallel-plate capacitor resonator. However, to find the loss tangent of the films in these devices, we assume the functional form for TLS loss in amorphous films \cite{VONSCHICKFUS1977}
\begin{align}
\tan\delta(\vec{r})=\frac{\tan\delta_0 \tanh(\hbar\omega/2k_BT)}{\sqrt{1+(E(\vec{r})/E_c)^2}},
\label{eq:tandForm}
\end{align}
and generate the field distribution $E(\vec{r})$ using a COMSOL field simulation. Then we use $\tan\delta_0$ and $E_c$ as fit parameters to match $1/Q_i$ in the simulation to the data (\textit{cf.} Sandberg \textit{et al.}\cite{Sandberg2012}). The intrinsic loss tangent is $\tan\delta_0=P_0p^2/3\epsilon$, where $p$ is the TLS dipole moment and $P_0$ is a materials constant defined by the TLS density $d^2n=P_0 d\Delta d\Delta_0/\Delta_0$, where $\Delta$ and $\Delta_0$ are the asymmetry energy and tunneling rate, respectively, in the standard model. \cite{Phillips1972,VONSCHICKFUS1977}

For the two amorphous films, a single power-dependent loss term representing the film thickness was insufficient to fit $Q_i$ and a second surface-loss mechanism was added and will be justified with impurity analysis below. The thickness of the lossy material at the surface is assumed to be 5 nm for the purposes of loss tangent analysis. The fit parameters for the curves of Fig.~\ref{fig:mat_loss} are shown in Table~\ref{tab:FitPar}. We find that the dominant loss occurs in the bulk of the crystalline film and the surface of the amorphous films. A relatively small loss term was also included in the fit and labeled as the floor in Table~\ref{tab:FitPar}. This power-independent loss is similar to the loss found in our reference resonators without the ALD dielectric, and is caused by an extrinsic effect unrelated to TLS loss such as the excitation of quasiparticles in the superconductor by infrared radiation. \cite{Barends2011}

The high thermal conductivity at low temperature \cite{Slack1971} and the low loss tangent near room temperature \cite{Daywitt1985} of crystalline BeO suggests that it \cite{Yum2012} may be a low-loss microwave dielectric for superconducting devices. The agreement of the $\rm{c-BeO}$ loss tangent to Eq.~\ref{eq:tandForm}, shown in Fig.~\ref{fig:mat_loss}, indicates that the TLS in this material have a nearly amorphous distribution of TLS such that $~1/|E|$ TLS saturation near resonance is observed. In contrast, a $~1/|E|^2$ dependence is expected for a crystal with identical defects. As a result, we find that TLS are described by an amorphous distribution in this single-phase oxide crystal. Previously an amorphous TLS distribution was observed in mixed alkali halide crystals,\cite{DeYoreo1986} but our result represents an unusual case in which TLS from a single-material crystalline film take on an amorphous distribution. In addition we see that the bulk loss tangent of $\rm{c-BeO}$ is approximately 6 times greater than that of the amorphous films, a value even higher than that found at room temperature in a pure crystal of the same material. \cite{Daywitt1985} Since BeO is a crystal, we expect the TLS nanostructure of this material to be more readily modeled and identified than in amorphous materials, which have generally unknown TLS structures. As seen in Table~\ref{tab:FitPar}, $\rm{a-LaAlO_3}$ shows a bulk loss tangent that is approximately equal to $\rm{a-Al_2O_3}$, suggesting that the TLS are related. The bulk loss of $\rm{a-Al_2O_3}$ is a factor of 2.3 lower than previous loss measurements of thermally grown $\rm{a-Al_2O_3}$. \cite{Martinis2005}

\begin{figure}
\includegraphics[trim = 0.1cm 0 0.2cm 0, clip]{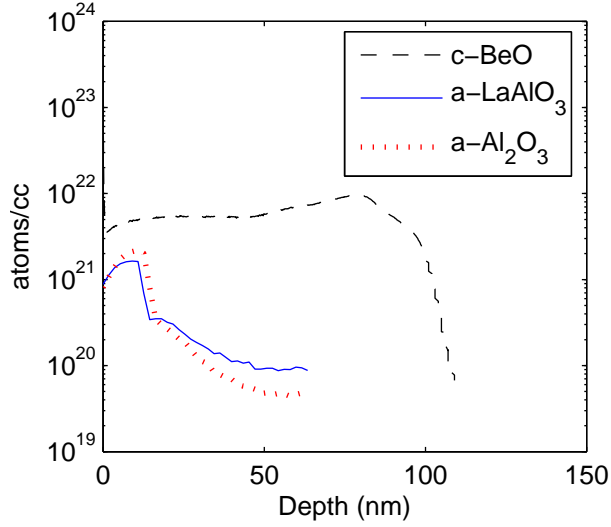}
\caption{\label{fig:mat_SIMS} Hydrogen impurity concentration measured by SIMS for films with nominally identical parameters to those shown in Fig.~\ref{fig:mat_loss}. The crystalline BeO film (75 nm thick) shows a large hydrogen concentration that is nearly uniform throughout the film depth, while the $\rm{Al_2O_3}$ and $\rm{LaAlO_3}$ films (each approximately 48 nm thick) show hydrogen concentrated at the surface of the films.}
\end{figure}

\begin{table}
\caption{\label{tab:FitPar} Fit parameters for the curves in Fig.~\ref{fig:mat_loss}. The bold values correspond to the dominant loss terms of the films, in the surface of the amorphous films and the bulk of the crystalline film. All films were fit assuming a bulk loss value, a floor limited by extrinsic effects, and amorphous films were fit with an additional 5 nm thickness of lossy material at the top surface.}
\begin{ruledtabular}
\begin{tabular}{llll}
\textbf{Fit parameters} &$\mathbf{c-BeO}$ &$\mathbf{a-Al_2O_3}$ &$\mathbf{a-LaAlO_3}$\\
\hline\\
$\tan\delta_{0, surface}(\times 10^{-3}$) &- &\textbf{13.0} &\textbf{4.5}\\
$E_{c, surface}$(V/m) &- &\textbf{0.75} &\textbf{0.55}\\
$\tan\delta_{0, bulk}(\times 10^{-3}$) &\textbf{6.2} &0.70 &1.1\\
$E_{c, bulk}$(V/m) &\textbf{0.70} &5.9 &60\\
$\tan\delta_{0, floor}(\times 10^{-6}$) &32 &4.0 &0.68\\
\end{tabular}
\end{ruledtabular}
\end{table}

While the nanostructure of these TLS has never been determined, it is believed that excess hydrogen can lead to rotating OH bonds that act as tunneling TLS in $\rm{a-SiO_2}$,\cite{Phillips1987} and $\rm{a-Al_2O_3}$.\cite{Holder2013} Hydrogen defects are also believed to limit room temperature electrical performance in  $\rm{a-Al_2O_3}$. \cite{Jennison2004} Secondary ion mass spectrometry (SIMS) measurements were performed on these films to determine the impurities present that could act as TLS defects. The hydrogen concentration is plotted for all three film types as a function of film depth in Fig.~\ref{fig:mat_SIMS}, and shows that there is an 8 times greater concentration in the bulk of the $\rm{c-BeO}$ film than in the amorphous films. In the BeO films the hydrogen is almost uniformly distributed throughout the bulk of the film (although some H diffuses into the Si substrate) which explains why only a bulk loss term is required in the fitting of this film. In contrast, the amorphous films have a large peak in hydrogen concentration at the surface, which explains why an additional surface loss term is required. This distribution of hydrogen shows that $\rm{c-BeO}$ has incorporated hydrogen during growth with its precursors of dimethyl beryllium and water, while the amorphous oxides have mostly incorporated hydrogen from ambient exposure rather than growth. The correlation of low temperature TLS loss with excess hydrogen strongly suggests that TLS are caused by hydrogen impurities. Recent simulations find that the nanostructure of TLS in $\rm{a-Al_2O_3}$ is consistent with hydrogenated cation vacancy defects.\cite{Holder2013}

\begin{figure}
\includegraphics[trim = 0cm 1.5cm 0.2cm 0, clip]{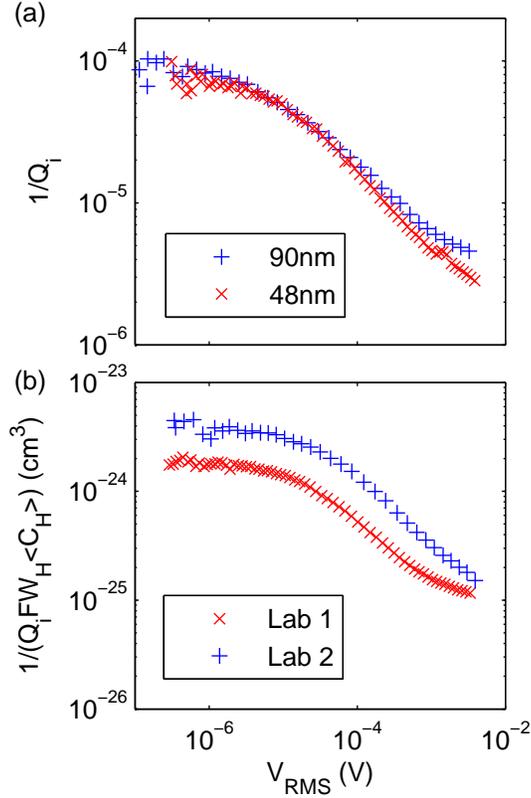}
\caption{\label{fig:thickness_loss} (a) Inverse internal quality factor of the resonators measured on nominally the same films grown in the same chamber to two different thicknesses. Because the loss is primarily at the surface, the quality factor is identical and does not scale with the thickness of the film. (b)	Ratio of the inverse internal quality factor (Eq.~\ref{eq:tand_avg}) to the product of the average hydrogen impurity concentration, the filling factor, and the weighting coefficient $(\langle C_x\rangle FW_x)$ of two $\rm{a-Al_2O_3}$ films grown in different laboratories. At low fields this reduces to the proportionality constant of hydrogen impurities to loss tangent, $K_H$.}
\end{figure}

As seen from Fig.~\ref{fig:mat_SIMS} most of the impurities in the $\rm{a-Al_2O_3}$ films are at the ambient-exposed surface. To verify that the $\rm{a-Al_2O_3}$ loss tangent is dominated by surface defects, we measured two films that are nominally identical in every way except for their thicknesses (48 and 90 nm). Figure \ref{fig:thickness_loss}(a) shows the inverse $Q_i$ measurement of the resonators on two identical films with differing thicknesses measured at 34 mK. If the TLS were uniformly distributed in the dielectric, then the $Q_i$ of the resonator would differ by a factor of 1.7, the difference in electrical energy stored in the film. Instead, the quality factor of the resonators matches to within 15\% at the low-voltage limit, confirming that the loss is primarily at the surface. We therefore believe that the development of in-situ deposition of dielectrics within MIM trilayer structures would greatly reduce dielectric loss.\cite{Kozen2013} In addition there is a possibility of growing ALD dielectrics without H-containing precursors.

\begin{figure}
\includegraphics[trim = 0cm 0 0.2cm 0, clip]{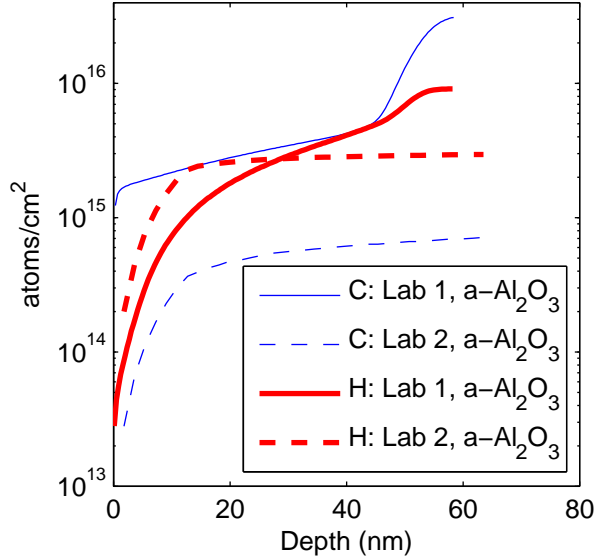}
\caption{\label{fig:thickness_SIMS} The integrated hydrogen and carbon concentration from the surface (0 nm) downward for aluminum oxide films from different labs as measured by SIMS. They show similar hydrogen concentration but the film from Lab 1 shows a much greater carbon concentration. Since the two films have similar loss and hydrogen concentrations, the results are consistent with H-based TLS.}
\end{figure}

One of the most common impurities in ALD films is carbon due to incomplete reaction of the organometallic precursors, in this case trimethylaluminum. We performed SIMS measurements for both carbon and hydrogen of $\rm{a-Al_2O_3}$ films grown in separate ALD chambers but with the same precursors. Figure~\ref{fig:thickness_SIMS} shows the integration of hydrogen and carbon impurities over the thickness of the films. The Lab 1 film has a much higher (40 times greater) carbon defect concentration than that of Lab 2. We measured resonators on these films and found them to have very similar $Q_i$'s, indicating that carbon plays a negligible role in the TLS loss. From Fig.~\ref{fig:thickness_SIMS} we see that hydrogen is a much more viable TLS candidate. To quantitatively analyze H and C impurities as TLS candidates we assume that the loss tangent,
\begin{align}
\tan\delta_{0,x}(\vec{r})=K_xC_x(\vec{r}),
\label{eq:tand_con}
\end{align}
is proportional to the impurity concentration $C_x(\vec{r})$ of species $x$, where $K_x$ is a TLS-loss proportionality constant. The low-voltage-amplitude internal quality factor of the film is then
\begin{align}
\frac{1}{Q_i(V\rightarrow0)}=K_x\langle C_x\rangle FW_x,
\label{eq:tand_avg}
\end{align}
related to $K_x$ by the average impurity concentration $\langle C_x \rangle=\int_{film}d^3r C_x(\vec{r})/(\int_{film}d^3r)$, the geometric filling factor, $F=\int_{film}d^3r \epsilon(\vec{r}) |E(\vec{r})|^2/\int_{all}d^3r \epsilon(\vec{r}) |E(\vec{r})|^2$, and the impurity weighting coefficient, $W_x=\int_{film}d^3r \epsilon(\vec{r}) |E(\vec{r})|^2C_x(\vec{r})/(\langle C_x \rangle\int_{film}d^3r \epsilon(\vec{r}) |E(\vec{r})|^2)$. Note that $F=1$ for a device with a single dielectric and $W_x=1$ for a uniform impurity distribution. In Fig.~\ref{fig:thickness_loss}(b) we plot $1/(Q_iFW_H\langle C_H \rangle)$ for the two films as a function of the microwave voltage using the measured values of $Q_i$ (at 80 mK) and $\langle C_H \rangle$, as well as the simulated field distributions for $W_H$ and $F$. At low field values, we find that the two films from different labs have $K_H~= 3 \times 10^{-24} cm^3$ to within 50\%. We consider the agreement between the two films to be reasonable since the films' hydrogen impurity concentration near the surface may have changed after the films were grown. In contrast, a similar analysis for carbon impurities and $K_C$ shows a difference of a factor of 33 between the two films, indicating that carbon is unrelated to TLS in our experimental conditions. This provides evidence that oxide films can be optimized for low temperature devices by lowering their hydrogen concentration.

In conclusion, we measured the loss of several ALD grown dielectric oxide films at millikelvin temperatures using superconducting resonators. From the voltage dependence of the loss we find that $\rm{c-BeO}$ exhibits an amorphous distribution of TLS. In addition, we find that the TLS loss of the crystalline film, $\rm{c-BeO}$, was higher than that of the amorphous films, $\rm{a-Al_2O_3}$ and $\rm{a-LaAlO_3}$, despite the fact that $\rm{c-BeO}$ can be a low-loss material at room temperature with high thermal conductivity at low temperatures. Using SIMS we correlated the low temperature loss with excess hydrogen defects on the surface of the amorphous films and in the bulk of the crystalline film. In $\rm{a-Al_2O_3}$ and $\rm{a-LaAlO_3}$ films, the bulk loss tangent was found to be similar, but the surface loss dominated such that the loss tangent was limited by hydrogen impurities. A thickness study of $\rm{a-Al_2O_3}$ films confirmed that the majority of the loss was on the surface. By testing $\rm{a-Al_2O_3}$ with different carbon concentrations, we found that carbon has a negligible effect on TLS loss. We conclude that various low temperature oxides can be optimized via rapid room temperature measurements of hydrogen.

\bibliography{ALD_paper}

\end{document}